\documentstyle[twocolumn,aps]{revtex}
\begin{document}
\draft 
\preprint{June 30, 1998}
\title
{
Collapse of Charge Gap
in Random Mott Insulators
}
\author{Y. Otsuka, Y. Morita and Y. Hatsugai}
\date {June 30, 1998}
\address
{
Department of Applied Physics, University of Tokyo,
7-3-1 Hongo Bunkyo-ku, Tokyo 113, Japan
}
\maketitle
\begin{abstract}
Effects of randomness on
interacting fermionic systems
in one dimension
are investigated 
by quantum Monte-Carlo techniques.
At first,
interacting spinless fermions are studied
whose ground state shows charge ordering.
Quantum phase transition due to randomness
is observed associated 
with the collapse of the charge ordering.
We also treat random Hubbard model
focusing on the Mott gap.
Although the randomness closes the Mott gap
and low-lying states are created,
which is observed in the charge compressibility,
no (quasi-) Fermi surface singularity is formed.
It implies localized nature of the low-lying states.
\end{abstract}


\narrowtext
Effects of Coulomb interaction and randomness
in fermionic systems
have been studied separately for a long time.
Their combined effects, however, remain unsolved 
especially in the strong-coupling regime.
The strong Coulomb interaction 
can cause the Mott transition and
lead the metallic system to the Mott insulator.
Also the randomness can make the electrons localized, 
which is the Anderson localization.
Although both are metal-insulator transitions, 
the physical characters are quite different.
For example, in the Mott insulator, 
the ground state has a finite charge gap 
due to the Coulomb interaction.
On the other hand, 
the insulator due to the Anderson localization
does not necessarily have a charge gap.
Therefore
the competition 
between the Coulomb interaction and the randomness 
is interesting from theoretical point of view
\cite{ref-rev3,ref-rev4,ref-rev1,ref-rev2}
and 
it may shed new light on some real materials.
In this paper, 
we systematically investigate 
one-dimensional fermionic systems 
with both Coulomb interaction and randomness
by quantum Monte Carlo (QMC) techniques
\cite{ref-wl1,ref-ax1,ref-ax2}.
Although the conventional wisdom is that
there is no quantum phase transition 
driven by randomness
in one and two dimensions \cite{ref-aalr},
we confirmed the existence of quantum phase transitions
due to combined effect 
of Coulomb interaction and randomness.
In this paper, at first,
we focus on the charge density wave (CDW) ordering
in interacting spinless fermions with randomness.
Next,
in order to understand low-lying excitations
of random Mott insulator, 
we study the charge compressibility
of random Hubbard model in the half-filled sector.
Also 
the one-particle Green's function is investigated to discuss
nature of the low-lying excitations.

{\it Random Spinless Fermions-}
To begin with,
we focus on CDW structure factor 
${\cal O}_{\rm CDW}$ of 
interacting spinless fermions 
with randomness in one dimension.
The Hamiltonian is given by
\begin{eqnarray}
{\cal H}_{sf} &=& {\cal H}_{v}+{\cal H}_{w}, \\
{\cal H}_{v }  &=& -t\sum _{\langle i,j \rangle}
(c_{i}^{\dagger}c_{j}+c_{j}^{\dagger}c_{i})
+V\sum _{\langle i,j \rangle}n_{i}n_{j}, \\
{\cal H}_{w }  &=& \sum _{i}w_{i}n_{i},
\end{eqnarray}
where $c_{i}^{\dagger}(c_{i})$
creates(annihilates) a spinless fermion at the $i$-th site,
$n_{i}=c_{i}^{\dagger}c_{i}$
and $\langle i,j \rangle$ is a nearest-neighbor link.
${\cal H}_{w}$ denotes site randomness
where $w_{i}$'s take $W$ or $-W$ at random.
We set $t=1$ as an energy unit and 
use open boundary condition.
In the following,
the system is set to be half-filled i.e. $N_{e}=L/2$,
where $N_{e}$ is number of the electrons
and $L$ is the system size.
The pure system without randomness is one of the typical
examples which show quantum critical phenomena.
When the interaction is weak ( $0{\le}V/t{\le}2$ ),
the system is metallic and
it belongs to
a universality class of the Tomonaga-Luttinger liquid.
A metal-insulator transition 
accompanied with a charge ordering
occurs at $V/t=2$.
For strong interaction ($V/t>2$),
the charge degree of freedom becomes frozen 
and the energy gap opens.
We focus on the charge-ordered phase ($V/t>2$)
where the charge gap is finite due to 
the nearest-neighbor interaction,
and study the effect of randomness.

In order to study the charge ordering 
when the randomness is included,
we calculate 
the CDW structure factor with momentum $\pi$ defined by
\begin{eqnarray}
{\cal O}_{\rm CDW}(T)=
\frac{1}{{\sharp}{\cal B}}
\sum_{i,j\in {\cal B}}
(-1)^{|i-j|}
{\langle (n_{i}-1/2)(n_{j}-1/2) \rangle},
\end{eqnarray}
where 
${\cal B}$ is set to be a subset of the lattice
to avoid a boundary effect,
${\cal B}=[L/4,3L/4]$ and 
${\sharp}{\cal B}$ is a number of sites in ${\cal B}$.
Without randomness,
${\cal O}_{\rm CDW}$ is strongly enhanced and diverging 
toward zero temperature
which corresponds to charge ordering in the ground state
due to Coulomb interaction ($V/t>2$).
On the other hand, 
when sufficiently strong randomness is included,
the charge ordering is expected to collapse and 
the diverging behavior of ${\cal O}_{\rm CDW}$ to vanish.
In order to obtain the ${\cal O}_{\rm CDW}$
for each realization of randomness,
world-line QMC method is employed \cite{ref-wl1}.
In Fig.1, temperature dependence of 
the ${\cal O}_{\rm CDW}$ is shown
for several strength of randomness
where simulations are performed 
in the half-filled sector with
typical system size $L=128$ and $V/t=3$.
Without randomness,
since the charge ordering occurs at zero temperature,
the structure factor shows diverging behavior 
toward zero temperature.
For weak randomness,
the structure factor 
still has similar behavior
down to the temperature we studied.
It means
(quasi-) long-range order
in the ground state 
or long localization length 
beyond the available system size.
On the other hand, 
when sufficiently strong randomness is included,
the temperature dependence of the ${\cal O}_{\rm CDW}$
shows qualitatively different behavior.
The rapid enhancement at low temperature vanishes.
It implies that the charge ordering 
completely fades out due to randomness.
In order to investigate detailed nature of the transition,
we study how the  structure factor depends 
on the randomness at $T=0.2t$.
As shown in the inset of Fig.1,
sudden decrease of ${\cal O}_{\rm CDW}$
at finite strength of randomness is observed.
It implies that the collapse of the charge gap occurs
at finite strength of randomness,
which is the order of the charge gap,
although we can not exclude a possibility that
the transition is the Kosterlitz-Thouless (KT) type
at infinitesimally weak randomness.

{\it Random Hubbard model-}
Next, in order to study effect of randomness 
on the Mott insulator, 
let us consider half-filled sector of
one-dimensional random Hubbard model.
The Hamiltonian 
is given by
\begin{eqnarray}
{\cal H}     &=& {\cal H}_{u}+{\cal H}_{w},   \\
{\cal H}_{u} &=&
 -t \sum _{\langle i,j \rangle \sigma}
           (c_{i \sigma }^{\dagger}c_{j \sigma}+ 
            c_{j \sigma}^{\dagger}c_{i \sigma})
 +U\sum _{i} n_{i \uparrow} n_{i \downarrow}, \\
{\cal H}_{w} &=& \sum _{i \sigma} w_{i} n_{i \sigma},
\end{eqnarray}
where $t$ is 
the nearest-neighbor hopping amplitude and
$U$ is the on-site Coulomb interaction.
${\cal H}_{w}$ denotes random potentials and
$w_{i}$'s are taken at random from the interval $[-w, w]$.
We treat the system in a grand canonical ensemble
with the chemical potential $\mu$. 
The boundary condition is periodic.
In the absence of randomness,
an infinitesimal interaction $U$ causes
a charge gap (Mott gap) $E_g$ at half filling ($\mu = U/2$).
The charge gap $E_g$ is exponentially small
in the weak-coupling region ($U/t \ll 1$) and 
linear in $U$ in the strong-coupling region ($U/t \gg 1$), for example,
estimated $E_g \simeq 1.3t$ for $U/t = 4$ \cite{ref-lw}.  
Here we shall discuss effect of randomness
on the Mott insulator
(see also Ref.\cite{ref-fk}).

To obtain 
approximation-free results, 
we use
a finite-temperature auxiliary-field QMC method \cite{ref-ax1,ref-ax2}.
Since we use the grand canonical ensemble,
there is a finite charge fluctuation
which is crucial 
for the knowledge of low-lying excitations.
Although the random potential 
breaks the particle-hole symmetry
for each realization of randomness,
half-filling condition is recovered 
after averaging over different realizations of randomness.
Our simulations are performed 
with system size $L = 36$ at $U/t=4$.
Severe finite-size effect 
due to energy discretization is observed 
in low-temperature region (lower than $T\sim 0.2t$ for $L=36$)
and the data for that region are not shown.
Moreover, 
since the particle-hole symmetry is broken for 
each realization of randomness,
negative-sign problem occurs in general.
However, in the parameter region 
we investigated,
it is not serious and 
the data are obtained
with sufficient accuracy.
To investigate 
the change of low-lying excitations
due to randomness,
we calculate charge compressibility 
$\kappa$ defined by

\begin{eqnarray}
\label{hub-kappa}
\kappa (T)= \frac{1}{L} \frac{\partial N_{e}}{\partial \mu} 
= \frac{\beta}{L}
\left
( {\langle \hat{N_{e}}^{2}\rangle} 
- {\langle \hat{N_{e}}    \rangle}^{2}
\right).
\end{eqnarray}

It measures 
fluctuation in the charge sector
and shows thermally-activated behavior
when the system has a finite charge gap.
On the other hand, in the absence of the charge gap, 
the charge compressibility 
is expected to be finite due to the low-lying excitations.
For example, in the non-interacting case,
$\kappa(T=0)$ is equal to the density of states 
at the Fermi energy.
Fig.~2 shows temperature dependence 
of the charge compressibility $\kappa$
for different strength of randomness.
Although a snap-shot for 
a particular realization of randomness
is shown in Fig.2,
the simulations for other realizations
were also performed
and we confirmed that
the global feature does not depend on
each realization.
Without randomness, 
$\kappa$ decreases toward zero 
and shows thermally-activated behavior
as the temperature is lowered 
indicating the existence of a charge gap.
This is a typical feature of the Mott insulator.
As randomness is turned on,
the enhancement in $\kappa$ is observed 
for all temperature.
For weak randomness, $w/t {\lesssim} 1.5$, 
although $\kappa$ is enhanced, 
it still seems to show thermally-activated behavior
as in the pure case.
On the other hand,
with strong randomness $w/t {\gtrsim} 1.5$,
there is no tendency for $\kappa$ to decrease
down to the lowest temperature we studied
and it seems that $\kappa$ is finite at $T=0$.
It suggests the collapse of the charge gap.
These behaviors imply that
sufficiently strong randomness,
which is the order of the charge gap,
takes the system away from the Mott insulator {\cite {ref-rev1,ref-rev2}}.
This is a quantum phase transition
of the Mott insulator driven by randomness.
Although it has been confirmed
that low-lying excitations are created in the Mott gap
with sufficiently strong randomness, 
the nature of the low-lying excitations is crucial
to understand the phase transition.
In order to study
the low-lying excitations, 
we calculate the one-particle Green's function
$G_{ij{\sigma}}
=\langle c_{i{}\sigma}^{\dagger}c_{{j}\sigma} \rangle$.
The feature of $G_{ij{\sigma}}$ in metallic states is 
clear in the momentum-space representation,
which is the momentum distribution function $n_{\sigma}(k)$.
Metallic nature of the system is reflected
by the singularity of $n_{\sigma}(k)$ at $k=k_F$.
For example,
it is a step function in the Fermi liquid (Fermi surface)
and, even in the Tomonaga-Luttinger liquid
(doped Hubbard chain),
the derivative of  $n_{\sigma}(k)$ is diverging at $k=k_F$
(quasi-Fermi surface).
On the other hand, 
in the Mott Insulator (half-filled Hubbard chain),
the derivative is finite.
We may summarize that 
the formation of (quasi-) Fermi surface, 
as temperature is lowered,
implies metallic nature of the system.
Fig.~3 shows momentum distribution function
$n_{\sigma}(k)$ at $T=0.2t$
for several randomness strength.
By performing calculations with different temperatures,
we confirmed that
the temperature is sufficiently low that 
the results reflect the ground-state properties.
Although
the charge gap is closed
for $w/t {\gtrsim} 1.5$ as discussed above,
the momentum distribution function is almost unchanged and
no singularity appears
at $k_{F}$ even for $w/t =2.0$.
It is in contrast to the transition
from the Mott insulator 
to the Luttinger liquid
in the pure Hubbard model
when the system is doped.
It suggests that the low-lying excitations 
created by randomness are localized.

In summary,
we have studied effects of randomness
in the interacting fermionic system
with charge gap (Mott gap).
The strong randomness closes the charge gap
and low-lying states are created, 
while (quasi-) Fermi surface is not formed.
It implies that the transition is 
an insulator to an insulator transition.
The transition point 
may be continuously connected to 
the metal-insulator transition point
in the pure system,
then it gives a new example of quantum phase transitions
in low-dimensional random fermionic systems.
These phenomena may give a clue for the understanding
of several aspects in
quantum phase transitions in low-dimensional random systems.

One of the authors (Y.M) 
is grateful to S.~Fujimoto for helpful correspondence.
Y.H is supported in part by  Grant-in-Aid
from the Ministry of Education, Science and Culture
of Japan and also by the Kawakami Memorial Foundation.
The computation in this work has been done
using the facilities of the Supercomputer Center,
ISSP, University of Tokyo.

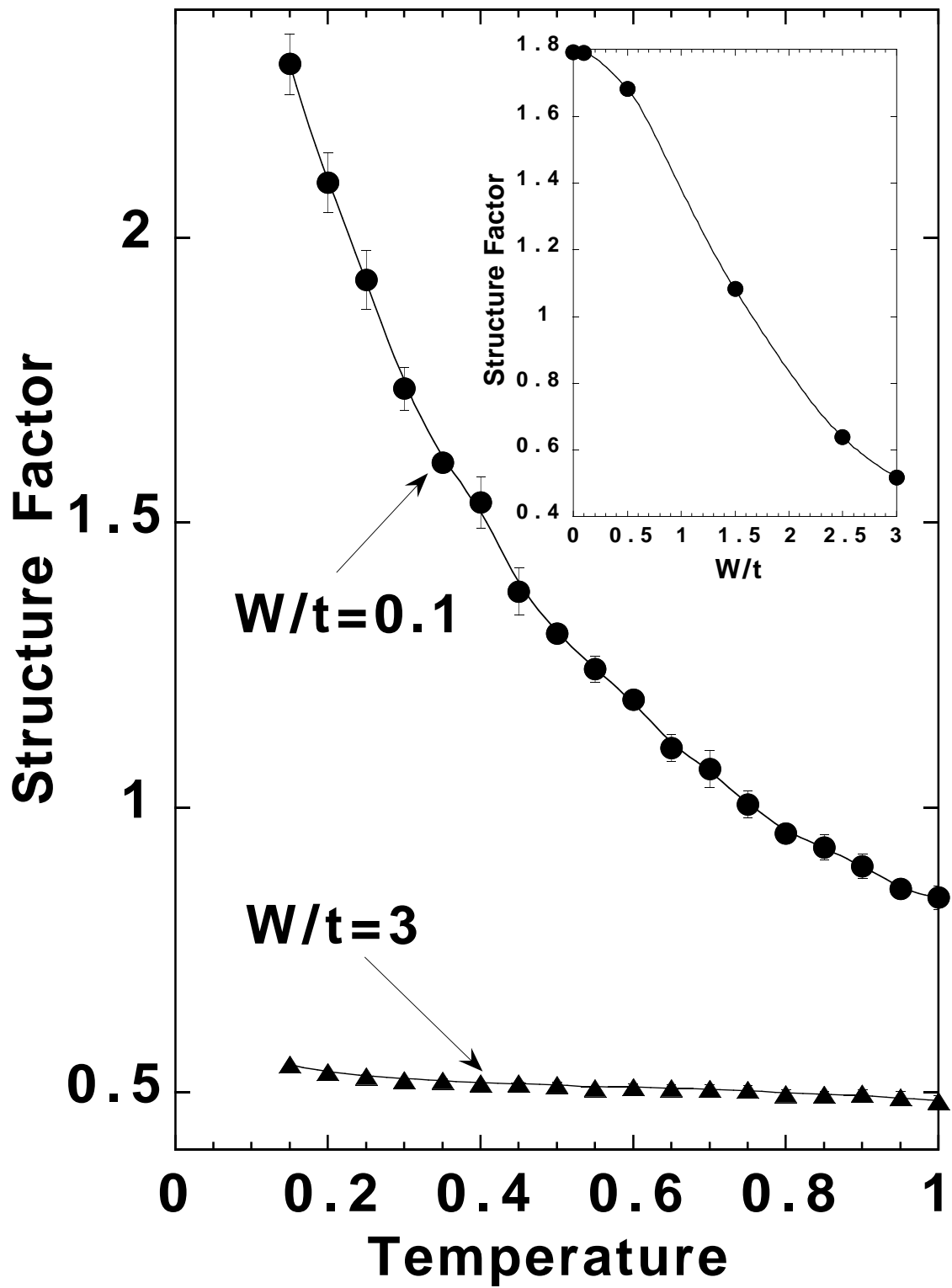
\begin{figure}
\caption{
The CDW structure factor 
as a function of temperature
for a particular realization of randomness 
($L=128$ and $V/t=3$).
For weak randomness ($W/t=0.1$),
the structure factor shows diverging behavior
down to the temperature we studied.
On the other hand, for strong randomness ($W/t=3$),
the divergence vanishes.
The inset shows 
how the structure factor 
at $T/t=0.2$
depends on randomness,
where $L=64$, $V/t=3$ and 
average over $30-40$ realizations of randomness
is performed. The line is guide for eyes.}
\end{figure}

\begin{figure}
\caption{
Temperature dependence of 
the charge compressibility $\kappa$ 
for a particular realization of randomness 
($L=36$ and $U/t=4$).
Without randomness, 
$\kappa$ shows 
thermally-activated behavior 
and decreases toward zero
at low temperature indicating
the existence of charge gap.
With weak randomness, $w < w_{c}$ $(w_{c}/t{\sim}1.5)$,
$\kappa$ still shows thermally-activated behavior. 
On the other hand, for strong randomness, $w > w_{c}$,
$\kappa$ does not decrease down to the temperature we studied.}
\end{figure}

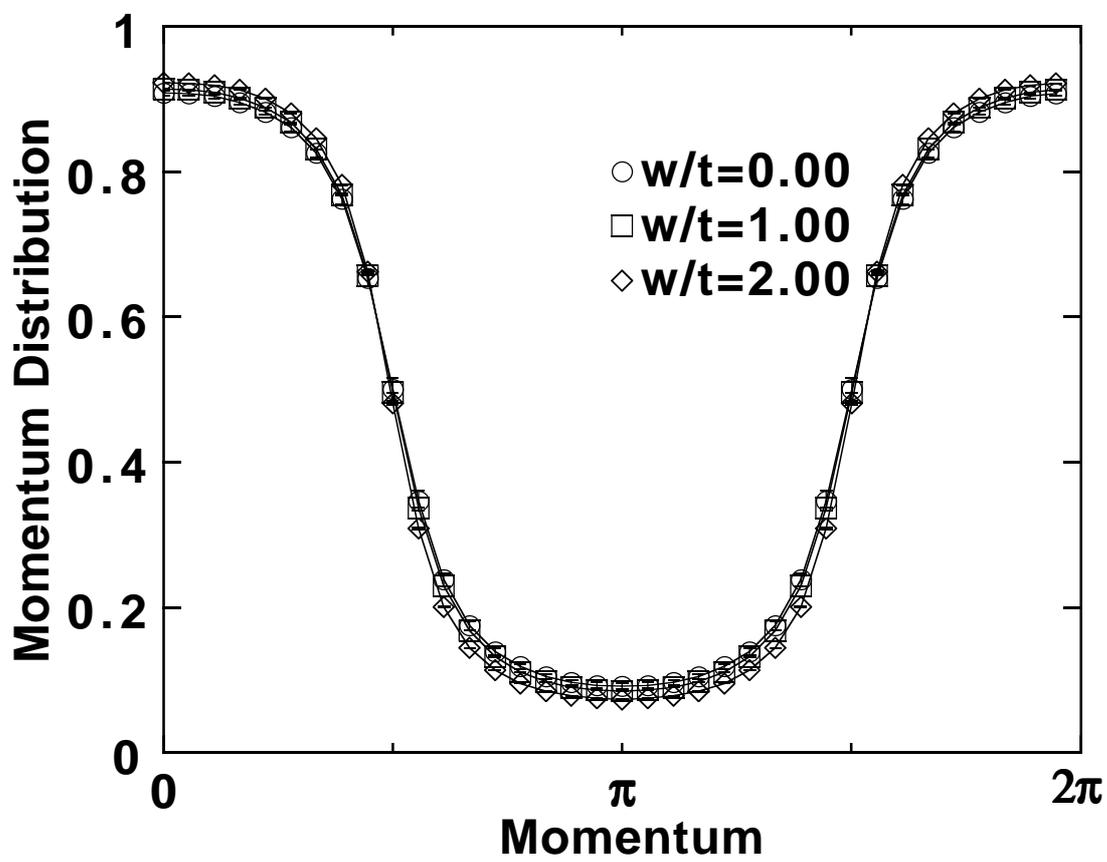
\begin{figure}
\caption{
The momentum distribution function
($L=36$ and $U/t=4$).
Average over about 10 realizations of randomness is performed.
There is no singularity 
at $k=k_{F}$
even for $w/t=2.0$ 
where the charge gap is closed 
as shown in the charge compressibility 
}
\end{figure}

\end{document}